\documentclass[aps,reprint,twocolumn,notitlepage,superscriptaddress,longbibliography]{revtex4}
\usepackage[T1]{fontenc}
\usepackage[utf8]{inputenc}
\usepackage{lmodern}
\usepackage{amsmath,amssymb,amsfonts,bm}
\usepackage{graphicx}
\usepackage{xcolor}
\usepackage[normalem]{ulem}
\usepackage{dcolumn}% Align table columns on decimal point
\usepackage{bm}% bold math
\usepackage{booktabs}
\usepackage[hidelinks]{hyperref}

\usepackage{multirow}
\usepackage{caption}
\usepackage{subcaption}  
\usepackage[mathlines]{lineno}
\usepackage{siunitx}

\begin{document}

\title{Rician distribution as a physically interpretable model for wind speed statistics}
\author{S. Mitra}
\affiliation{Nonlinear and Non-equilibrium Physics Unit, OIST Graduate University, 1919-1 Tancha, Onna, Okinawa, 904-0495, Japan}
\author{S. E. Lakhal}
\affiliation{Nonlinear and Non-equilibrium Physics Unit, OIST Graduate University, 1919-1 Tancha, Onna, Okinawa, 904-0495, Japan}
\affiliation{Center for Mathematical Morphology (CMM), MINES ParisTech, 35 rue Saint Honoré, 77305, Fontainebleau, France}
\author{C. P. Connaughton}
\affiliation{London Mathematical Laboratory, 8 Margravine Gardens, London W6 8RH, UK}
\author{J. E. Sardonia}
\affiliation{Exus Renewables North America, Pittsburgh, Pennsylvania, USA}
\author{M. M. Bandi}
\email[]{bandi@oist.jp}
\affiliation{Nonlinear and Non-equilibrium Physics Unit, OIST Graduate University, 1919-1 Tancha, Onna, Okinawa, 904-0495, Japan}

\date{\today}

\begin{abstract}

The statistics of atmospheric wind variations are commonly modeled using Gaussian or Weibull forms, which often trade physical interpretability against statistical accuracy, especially in the distribution tails. Here we derive a Rician distribution for wind speed from a simple physical model based on two orthogonal Gaussian velocity components with a non-zero mean in the preferred direction. Using wind-speed records from four geographically distinct wind farms, we show that the Rician model consistently outperforms the Gaussian model and remains competitive with the Weibull model. The same behavior persists when the data are partitioned into monthly windows, where the Rician parameters also provide a transparent description of seasonal and geographic variability, compared to Weibull parameters. In addition, the model naturally connects Gaussian-like and Weibull-like regimes through the Rician parameter ratio $\frac{\mu}{\sigma}$, making the Rician distribution a compact and physically interpretable two-parameter model for wind-speed statistics.

\end{abstract}
                              
\maketitle

%%%%%%%%%%%%%%%%%%%%Introduction%%%%%%%%%%%%%%%%
\section{Introduction}
\label{sec:intro}

Wind-speed modeling is central to a wide range of applications, including aviation safety \cite{POBOCIKOVA2017713,11369485}, weather forecasting \cite{9772879,yamaguchi}, renewable-energy systems \cite{apt2007spectrum,bel2016grid,bandi2017spectrum,bandi2016variability}, and environmental processes such as pollutant dispersion and wildfire spread \cite{MEI20161102,LIU2023103974}. Its importance has grown with the rapid expansion of renewable energy, where weather-driven variability directly impacts grid reliability and operational flexibility \cite{chum2011ipcc,apt2014variable,schmietendorf2017impact,smith2022effect}. In particular, wind and solar resources are governed by atmospheric dynamics that cannot be dispatched on demand \cite{mackay2016sustainable,apt2007spectrum,bel2016grid,bandi2017spectrum,bel2019geographic,bel2024spectral,bandi2016variability}. As their grid share rises, this variability affects adequacy and operational flexibility, including effective load-carrying capability (ELCC) \cite{6963441} and inertia-related frequency response \cite{ajami2026impact,bevrani2009robust,10413919,PADHAN2013242}.

Wind-speed is intrinsically multiscale and nonstationary \cite{mackay2016sustainable,apt2007spectrum,bel2016grid,lakhal2026collective}: while high frequency fluctuations already display the intermittent nature of atmospheric turbulence, diurnal, synoptic and seasonal cycles induce persistent and non-stationary shifts in the underlying statistics \cite{PhysRevE.110.041001,Carbone,vonbrandits}. Additionally, it is now well established that local topographic variations too play a role in wind speed patterns and their variability \cite{Jiang2008, Dvorak2010, bandi2016variability}. Consequently, fitting a single static distribution over long records can obscure physically meaningful temporal structure and weaken interpretation across sites and seasons.

The empirical probability density function (p.d.f) of wind-speed typically exhibits a well-defined bulk together with a heavy upper tail associated with strong wind events \cite{akdaug2010use,CARTA2009933,CARTA2007518} (Fig.~\ref{fig:monthly_hist}). A useful model should therefore capture both regimes while retaining parameters that remain physically interpretable across seasons and geographic locations. Canonical choices \cite{akdaug2010use,akdaug2009new,lencastre2024modeling,shi2021wind,WADI2023237,CELIK2006105,CARTA2009933,CARTA2007518,CELIK2003693,CHANG2011272,CHAURASIYA20182299,CHELLALI2012379} include Gaussian [Eq.~\ref{gaussian_pdf}] and Weibull [Eq.~\ref{Weibull_pdf}] distributions. Gaussian models describe fluctuations around a mean state but tend to underpredict extremes, whereas Weibull models better capture tail behavior but rely on phenomenological parameters \cite{akdaug2010use,akdaug2009new} that are less transparent for physical interpretation.

To address this trade-off, we model wind velocity as two orthogonal Gaussian components, with a non-zero mean in the  leading direction and zero mean in the perpendicular direction. Importantly, this leading direction is not fixed in earth referential, but can itself vary over time. Under this construction, the wind-speed magnitude follows a Rician distribution [Eq.~\ref{rician_pdf}] \cite{article}, which naturally accommodates heavier tails while it's parameters retain  clear physical interpretation.

This framework offers several potential advantages. First, the Rician distribution naturally emerges from a simple physical decomposition of the wind-velocity field and therefore provides a direct connection between model parameters and underlying flow characteristics. Second, unlike purely phenomenological descriptions, its parameters can be interpreted in terms of the coherent mean-flow component and the intensity of fluctuations around that mean. This raises the possibility of tracking temporal and spatial variability in wind conditions through physically meaningful quantities. Finally, the Rician distribution contains both Gaussian-like and Weibull-like behavior as limiting cases, suggesting that it may provide a flexible two-parameter framework capable of describing a broad range of wind regimes.

In the following sections, we investigate these hypotheses using wind-speed time series from four utility-scale wind farms located in distinct geographic regions. Key dataset characteristics are summarized in Table~\ref{Table:1}. Each turbine provides 10-minute averaged wind-speed measurements, following International Energy Commission standard 61400-12~\cite{IEC61400-12-1}, over approximately 2--5 years, enabling robust analysis of seasonal intra/inter-farm variability.

\begin{table}[ht]
    \small
    \centering
    \caption{Characteristics of the wind farms}
    \label{Table:1}
    \footnotesize
    \begin{tabular}{@{}cccc@{}}
    \toprule
    \textbf{Farm} & \textbf{Turbines ($N_{\mathrm{tur}}$)} & \textbf{Area (km$^2$)} & \textbf{Sampling time (years)} \\
    \midrule
    1       & 52                                     & 89.72                  & 2.7 \\
    2       & 80                                     & 68.80                  & 4.9 \\
    3       & 120                                    & 93.08                  & 3.6 \\
    4       & 64                                     & 80.94                  & 3.0 \\
    \bottomrule
    \end{tabular}
\end{table}

Our paper organizes as follows. Section~\ref{sec:motivation_new_model} motivates the need for a new wind-speed model. Section~\ref{sec:new_model} derives the Rician distribution from a physical framework. Section~\ref{sec:model_fitting} outlines the estimation procedure and comparison metrics. Section~\ref{sec:seasonal_geo_variability} analyzes seasonal and geographic variability, and Sec.~\ref{sec:model_comparisons} discusses limiting behavior.

%%%%%%%%%%%%%%%%%%%%%%%%%%%%%%%%%%%%%%%%%%%%%%Motivation for a new model for wind speed%%%%%%%%%%%%%%%%%%%%%%%%%%%%%%%%%
\section{Motivation for a New Wind-Speed Model}
\label{sec:motivation_new_model}

\begin{figure}[t!]
    \centering
\includegraphics[width=1.0\linewidth]{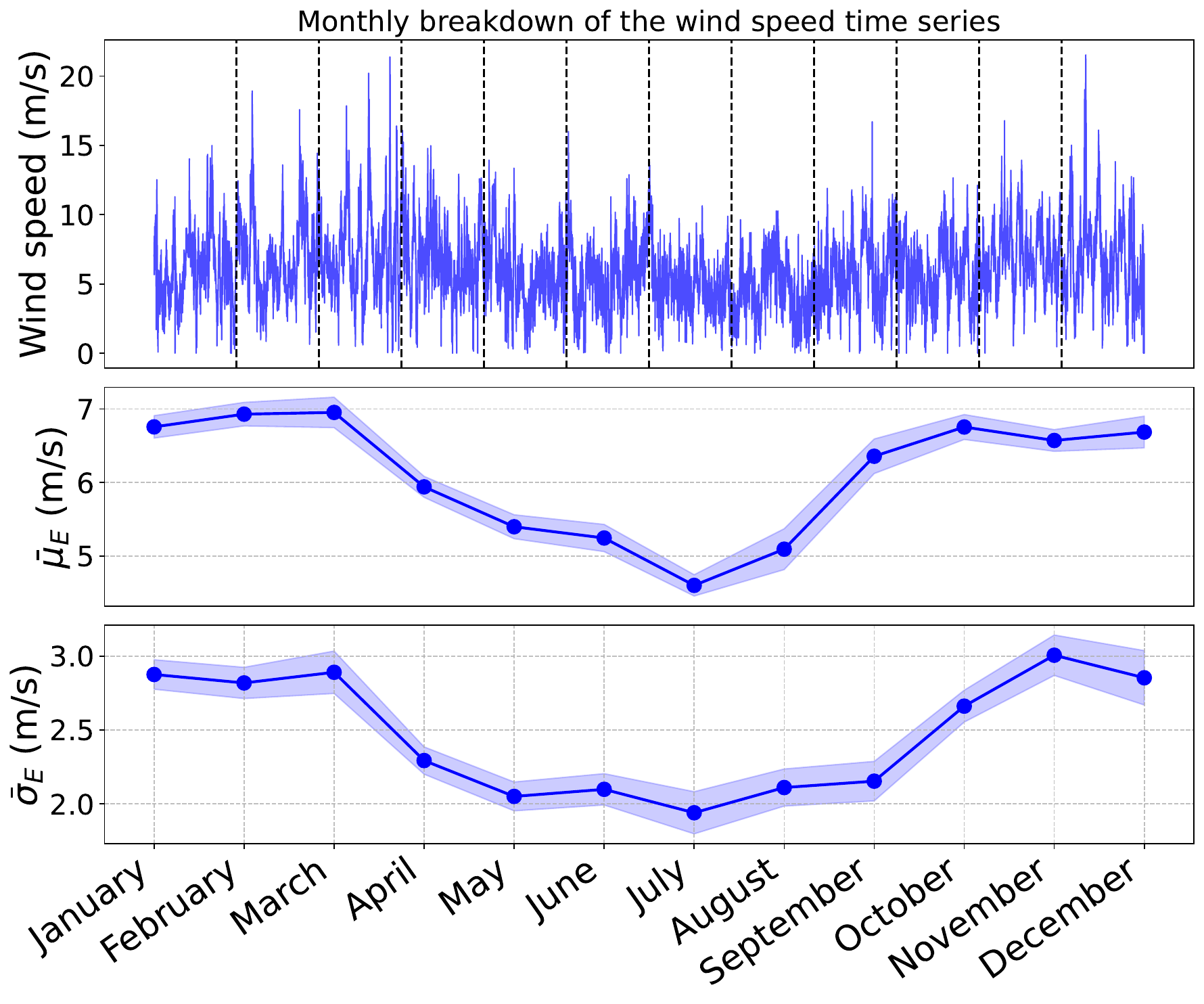}
\caption{Monthly decomposition of one year wind-speed data for a representative turbine in Wind Farm~1, together with farm-level ensemble statistics (mean and standard deviation [Eq.~\ref{farm_avg}]), averaged over all available years (see Table~\ref{Table:1}), by the solid lines and inter-turbine variability bands [Eq.~\ref{farm_std}], by the shaded color.}
\label{fig:monthly_timeseries}
\end{figure}

The nonstationary statistics of wind speed are a direct consequence of its multiscale spatial and temporal variability \cite{mackay2016sustainable,apt2007spectrum,bel2016grid,bandi2017spectrum,bandi2016variability}.
As a result, fitting a single distribution to an entire time series can obscure physically meaningful structure. We therefore partition each turbine time series into monthly windows and analyze statistics within each calendar month (Fig.~\ref{fig:monthly_timeseries}), providing a balance between sample size and stable weather conditions, where fluctuations are locally mean-reverting, while retaining seasonal modulation.

For each turbine and month, we compute the empirical mean and standard deviation, denoted $\mu_E$ and $\sigma_E$. To characterize farm-level behavior, we then construct ensemble statistics across turbines. For a given month $m$ and parameter $p \in \{\mu_E, \sigma_E\}$, the farm-level mean is
\begin{equation}\label{farm_avg}
\bar{p}_m = \frac{1}{N_{\mathrm{tur}}}\sum_{i=1}^{N_{\mathrm{tur}}} p_{i,m}
\end{equation}
and the corresponding inter-turbine variability is quantified by the sample standard deviation
\begin{equation}\label{farm_std}
s_{p,m} = \sqrt{\frac{1}{N_{\mathrm{tur}}-1}\sum_{i=1}^{N_{\mathrm{tur}}}\left(p_{i,m}-\bar{p}_m\right)^2}
\end{equation}
Here, $p_{i,m}$ denotes the value for turbine $i$ in month $m$, and $N_{\mathrm{tur}}$ is the number of turbines in the farm (Table~\ref{Table:1}).

%\begin{figure}
%\centering
%\includegraphics[width=\linewidth]{Fig1.pdf}
%\caption{Monthly decomposition of one year wind-speed data for a representative turbine in Wind Farm~1, together with farm-level ensemble statistics (mean and standard deviation [Eq.~\ref{farm_avg}]), averaged over all available years (see Table~\ref{Table:1}), by the solid lines and inter-turbine variability bands [Eq.~\ref{farm_std}], by the shaded color.}
%\label{fig:monthly_timeseries}
%\end{figure}

Figure~\ref{fig:monthly_timeseries} summarizes this decomposition. The farm-level means (solid lines) and variability bands $\bar{p}_m \pm s_{p,m}$ (shaded regions) reveal coherent seasonal trends alongside moderate inter-turbine variability. Since wind farms span several kilometers (Table~\ref{Table:1}), turbines are subject to common large-scale forcing but are susceptible to distinct local gust conditions, leading to variability across both space and time. Monthly aggregation therefore provides a practical quasi-stationary representation. In particular, although measurable differences exist among turbines within the same farm, these spatial variations are modest compared with the dominant seasonal modulation of the wind-speed statistics.

\begin{figure}[htb!]
\centering
\includegraphics[width=1\linewidth]{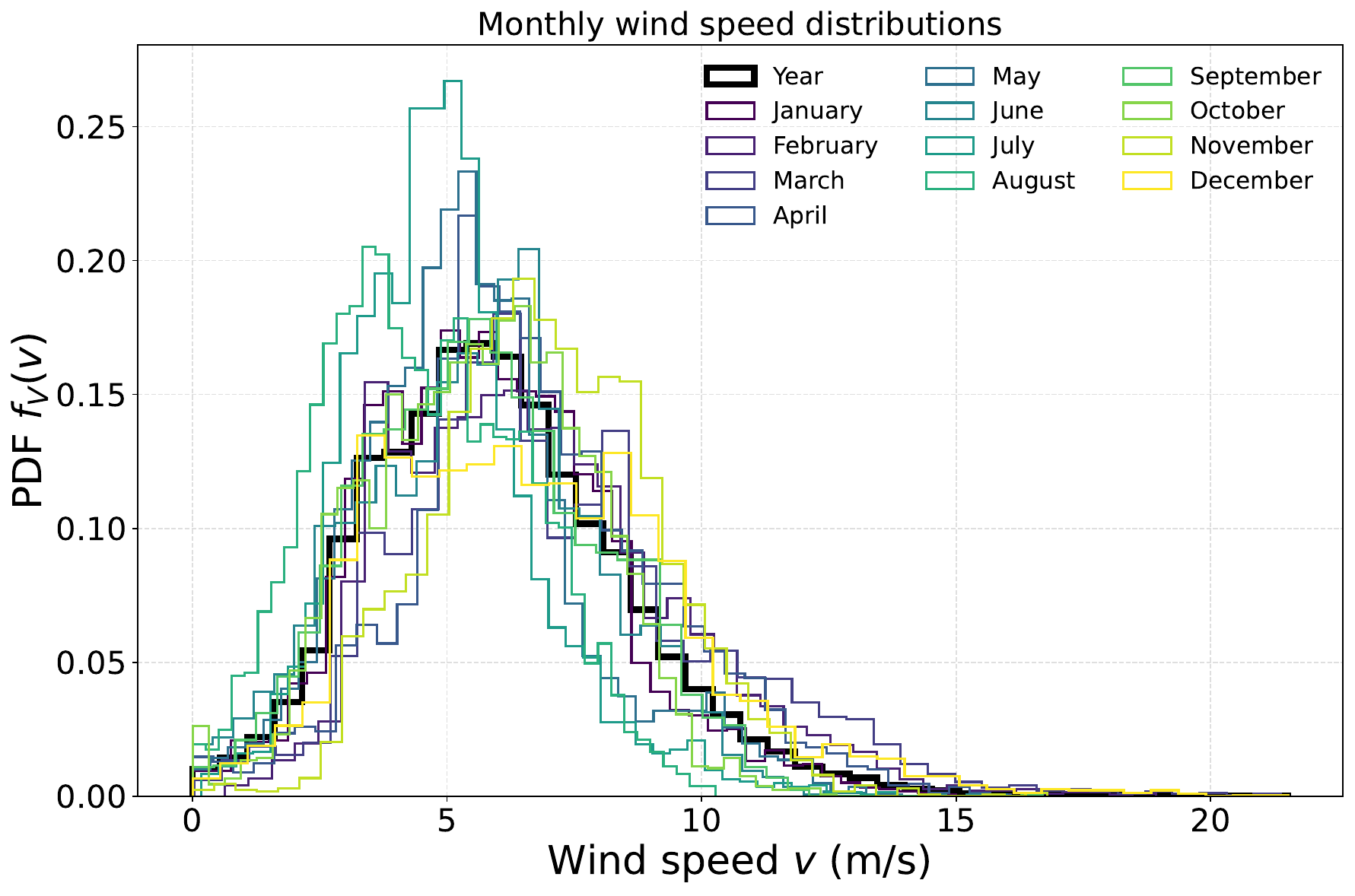}
\caption{Histograms of the one-year and monthly wind-speed records for the same turbine. Monthly decomposition  motivating empirical structure of wind-speed data: a stable bulk with a heavy upper tail with seasonal drift.}
\label{fig:monthly_hist}
\end{figure}

The corresponding wind speed distributions (yearly and monthly segmentation) (Fig.~\ref{fig:monthly_hist}) exhibit two robust features: (i) a seasonally varying bulk with evolving mean and fluctuations, and (ii) a pronounced upper tail associated with intermittent high-wind events. A suitable statistical model must capture both regimes simultaneously.

Previous studies \cite{akdaug2010use,akdaug2009new,lencastre2024modeling,shi2021wind,WADI2023237,CELIK2006105,CARTA2009933,CARTA2007518,CELIK2003693,CHANG2011272,CHAURASIYA20182299,CHELLALI2012379} typically employ either a Gaussian distribution,
\begin{equation}\label{gaussian_pdf}
    f_G(v;\zeta,\gamma)=\frac{1}{\sqrt{2\pi \gamma^2}}
    \exp\!\left[-\frac{(v-\zeta)^2}{2\gamma^2}\right]
\end{equation}
or a Weibull distribution,
\begin{equation}\label{Weibull_pdf}
    f_W(v;\beta,\eta)=
    \frac{\beta}{\eta}
    \left(\frac{v}{\eta}\right)^{\beta-1}
    \exp\!\left[-\left(\frac{v}{\eta}\right)^{\beta}\right]
\end{equation}
The Gaussian parameters $(\zeta,\gamma)$ directly represent the mean flow and fluctuation intensity, but the rapid decay of the distribution leads to systematic underestimation of high-wind events. In contrast, the Weibull distribution provides greater flexibility in capturing the upper tail, but its parameters $(\beta,\eta)$ lack a direct physical interpretation in terms of seasonal and geographic variability.

These limitations motivate the search for a model that simultaneously (i) captures extreme events, (ii) retains physically interpretable parameters linked to temporal and spatial intra and inter-farm variability, and (iii) reduces to Gaussian- or Weibull-like behavior in appropriate limits.

%%%%%%%%%%%%%%%%%%%%%%%%%%%%%%%%%%%%%%%%%%%%%%%%%%%%%%%%%%%%%%%%%%%%%%%%%%%%%%%%%%%%%%%%%%%%%%%%%%%%%%%%%%%%%%%
%%%%%%%%%%%%%%%%%%%%%%%%%%%%%%%%%%%%%%%%%%%%%The rician PDF derivation%%%%%%%%%%%%%%%%%%%%%%%%%%%%%%%%%%%%%%%%

\section{A refined wind speed distribution model \label{sec:new_model}}

To derive the wind-speed distribution, we model the instantaneous wind velocity $\vec{v}$ at a fixed point in space as a two-dimensional random vector, $\vec v = (v_\parallel,v_\perp)$, with one preferred direction. Specifically, we assume
$v_{\parallel} \sim \mathcal{N}(\mu,\sigma^2)$ and $v_{\perp} \sim \mathcal{N}(0,\sigma^2)$,
where $v_{\parallel}$ and $v_{\perp}$ are the parallel and perpendicular components of the wind velocity $\vec{v}$, respectively, and are treated as independent, and free from the Earth referential.

The wind speed, defined as the magnitude of the velocity, is

\begin{equation}
v = \sqrt{v_{\parallel}^2 + v_{\perp}^2}
\end{equation}

Under these assumptions, the resulting wind-speed distribution follows the Rician (Rice) form \cite{article} (a complete derivation is provided in \hyperref[appendix1]{Appendix~A}):

\begin{equation}\label{rician_pdf}
f_R(v;\mu,\sigma) = \frac{v}{\sigma^2}
\exp\left(-\frac{v^2+\mu^2}{2\sigma^2}\right)
I_0\left(\frac{v\mu}{\sigma^2}\right)
\end{equation}

The parameters of the distribution have a more direct physical interpretation than the Weibull parameters [Eq.~\ref{Weibull_pdf}] \cite{akdaug2010use,akdaug2009new}: $\mu$ characterizes the coherent mean-flow component, while $\sigma$ quantifies fluctuation intensity about that mean and are allowed to vary in time (e.g., across monthly windows), capturing seasonal and environmental changes in wind conditions (Fig.~\ref{fig:rician_params}). Compared with the Gaussian model [Eq.~\ref{gaussian_pdf}], the Rician distribution [Eq.~\ref{rician_pdf}] yields a heavier high-speed tail for large $v$ (Fig.~\ref{fig:ccdf_plot}).

\section{Model fitting and parameter estimation
\label{sec:model_fitting}}

We evaluate the proposed Rician model [Eq.~\ref{rician_pdf}] against the two canonical alternatives, Gaussian [Eq.~\ref{gaussian_pdf}] and Weibull [Eq.~\ref{Weibull_pdf}], using complete wind-speed records from one representative turbine in each of the four wind farms (see Table ~\ref{Table:1}). This full-record analysis provides the most stringent first benchmark because it tests whether a single model can capture the overall wind-speed statistics before any seasonal decomposition is introduced. Although the detailed wind conditions vary from turbine to turbine within a given farm (Fig.~\ref{fig:monthly_timeseries}), leading to changes in the fitted parameter values (Fig.~\ref{fig:rician_params}), the overall goodness-of-fit ranking remains qualitatively the same across turbines: Rician consistently performs better than Gaussian and remains comparable to Weibull. We then repeat the comparison on monthly windows to assess whether the same conclusions remain valid once the data are partitioned into subsets of stable weather conditions, hence with more stationary statistics. 

Model parameters are estimated by maximum likelihood estimation (MLE), which avoids histogram-dependent parameter bias and provides a standard likelihood-based comparison across models \cite{Newman_2005,Alstott_2014}. Specifically,

\begin{equation}\label{mle}
\hat{\theta} = \arg\min_{\theta} \left( -\sum_{i=1}^{N} \log f_V(v_i \mid \theta) \right)
\end{equation}

where $f_V(v_i \mid \theta)$ is the model PDF evaluated at observation $v_i$, and $N$ is the number of samples. The fitted parameters for all models and all wind farms are listed in Table~\ref{tab:wind_farm_comparison}.

\begin{table}[htbp]
\centering
\caption{Comparison of Distribution Fits Across Four Wind Farms}
\label{tab:wind_farm_comparison}

\resizebox{\columnwidth}{!}{%
\begin{tabular}{cccccc}
\toprule
\textbf{Farm} & \textbf{Model} & \textbf{Params.} & \textbf{KL} & \textbf{KS} & \textbf{Tail-$L_2$} \\ 
\midrule
\multirow{3}{*}{1} & Gaussian ($m/s, m/s$) & (6.28, 2.62) & 0.029 & 0.037 & 1.921 \\
                   & Weibull ($m/s, -$)    & (7.07, 2.53) & \textbf{0.024} & \textbf{0.026} & \textbf{0.731} \\
                   & Rician ($m/s, m/s$)   & (5.44, 2.90) & \textbf{0.024} & 0.027 & 1.076 \\
\midrule
\multirow{3}{*}{2} & Gaussian ($m/s, m/s$) & (8.13, 3.66) & 0.016 & 0.035 & 0.067 \\
                   & Weibull ($m/s, -$)    & (9.15, 2.32) & 0.017 & \textbf{0.016} & \textbf{0.004} \\
                   & Rician ($m/s, m/s$)   & (6.62, 4.23) & \textbf{0.015} & 0.019 & 0.006 \\

\midrule
\multirow{3}{*}{3} & Gaussian ($m/s, m/s$) & (8.06,3.24) & 0.013 & 0.028 & 0.096 \\
                   & Weibull ($m/s, -$)    & (9.07,2.67) & \textbf{0.005} & \textbf{0.007} & 0.047 \\
                   & Rician ($m/s, m/s$)   & (7.09,3.55) & 0.007 & 0.014 & \textbf{0.043} \\

\midrule
\multirow{3}{*}{4} & Gaussian ($m/s, m/s$) & (7.03,3.17) & 0.026 & 0.040 & 0.167 \\
                   & Weibull ($m/s, -$)    & (7.94,2.37) & \textbf{0.008} & \textbf{0.014} & \textbf{0.026} \\
                   & Rician ($m/s, m/s$)   & (5.65,3.71) &  0.012 & 0.023 & 0.031 \\

\bottomrule
\end{tabular}
}
\end{table}

To assess model quality across different parts of the distribution, we use three complementary representations \cite{cohen,book_majumdar,book_bala}: the probability density function (PDF) emphasizes the bulk structure, the cumulative distribution function (CDF) [Eq.~\ref{CDF_speed}] is particularly sensitive to cumulative discrepancies at low and moderate speeds, and the complementary cumulative distribution function (CCDF) [Eq.~\ref{CCDF_speed}] isolates the high-wind tail. Correspondingly, we use two global goodness-of-fit measures, the Kullback--Leibler (KL) divergence [Eq.~\ref{kl_div}] and the Kolmogorov--Smirnov (KS) distance [Eq.~\ref{ks}], together with a tail-restricted logarithmic $L_2$ metric [Eq.~\ref{l2_dist_tail}] designed to probe high-wind extremes more directly.

To quantify goodness of fit in the PDF representation, we use the discrete Kullback--Leibler (KL) divergence between empirical and model histograms \cite{Brunton_Kutz_2022}:

\begin{equation}\label{kl_div}
S(f_k,g_k) = \sum_{k} f_{k} \log \left(\frac{f_{k}}{g_{k}}\right)
\end{equation}

Here, $f_k$ and $g_k$ denote empirical and model probabilities in bin $k$, respectively. From an information-theoretic perspective, $S(f_k,g_k)$ measures the information loss incurred when $g_k$ is used to represent $f_k$.

\begin{figure}[htbp]
     \centering
     % --- PDF Plot ---
     \begin{subfigure}[b]{0.32\textwidth}
         \centering
         \includegraphics[width=\textwidth]{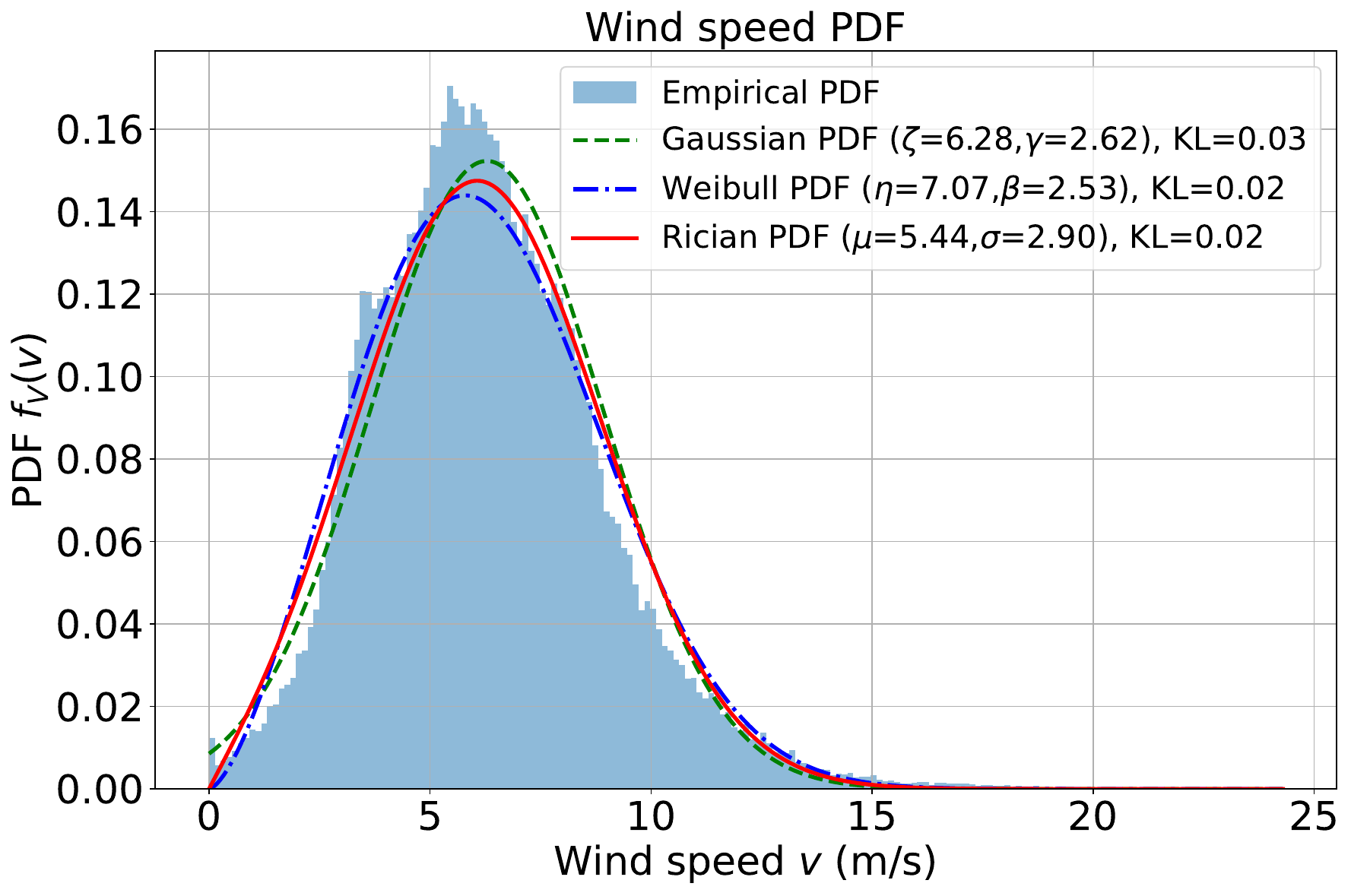}
         \caption{}
         \label{fig:pdf_plot}
     \end{subfigure}
     \hfill
     % --- CDF Plot ---
     \begin{subfigure}[b]{0.32\textwidth}
         \centering
         \includegraphics[width=\textwidth]{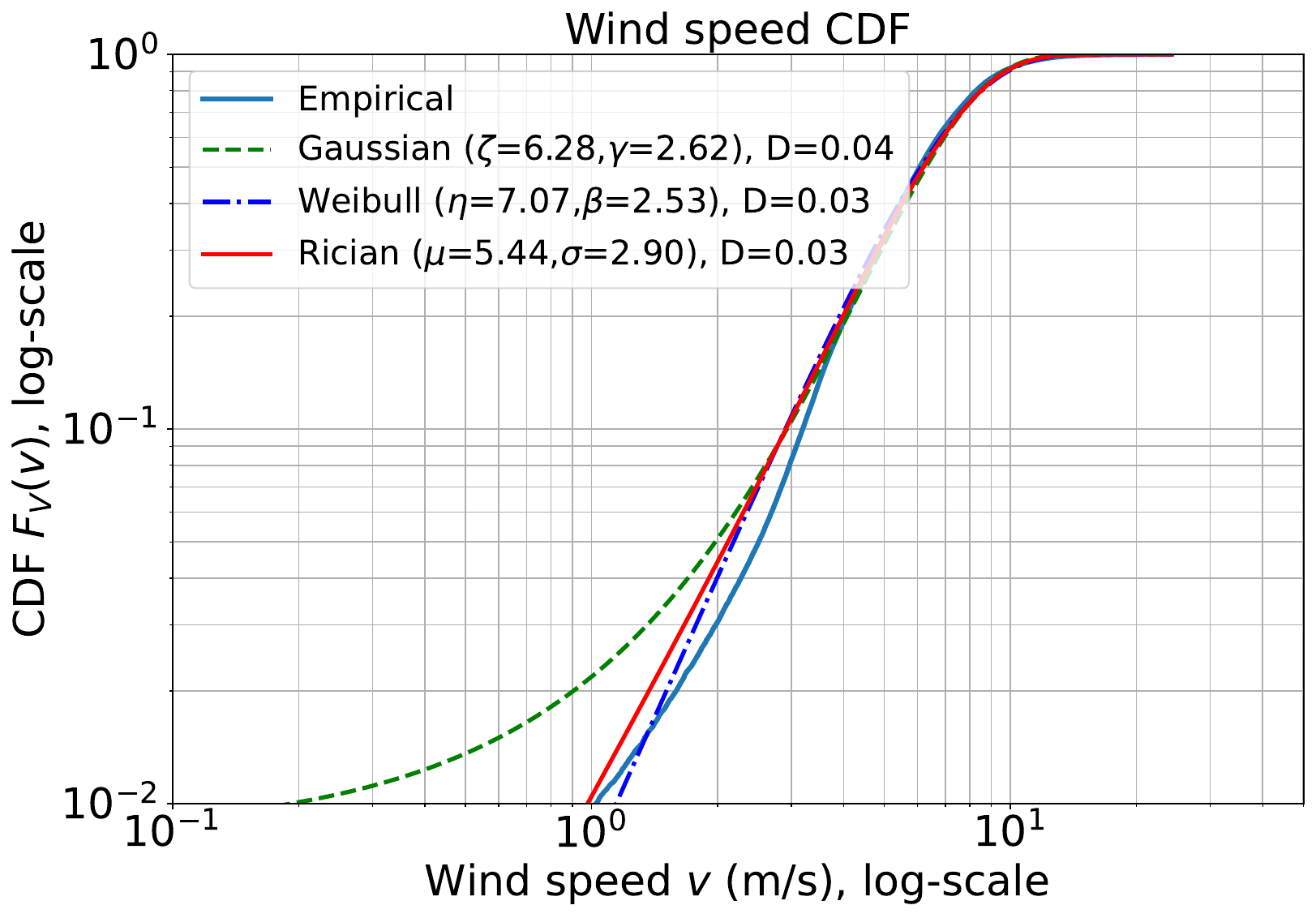}
         \caption{}
         \label{fig:cdf_plot}
     \end{subfigure}
     \hfill
     % --- CCDF Plot ---
     \begin{subfigure}[b]{0.32\textwidth}
         \centering
         \includegraphics[width=\textwidth]{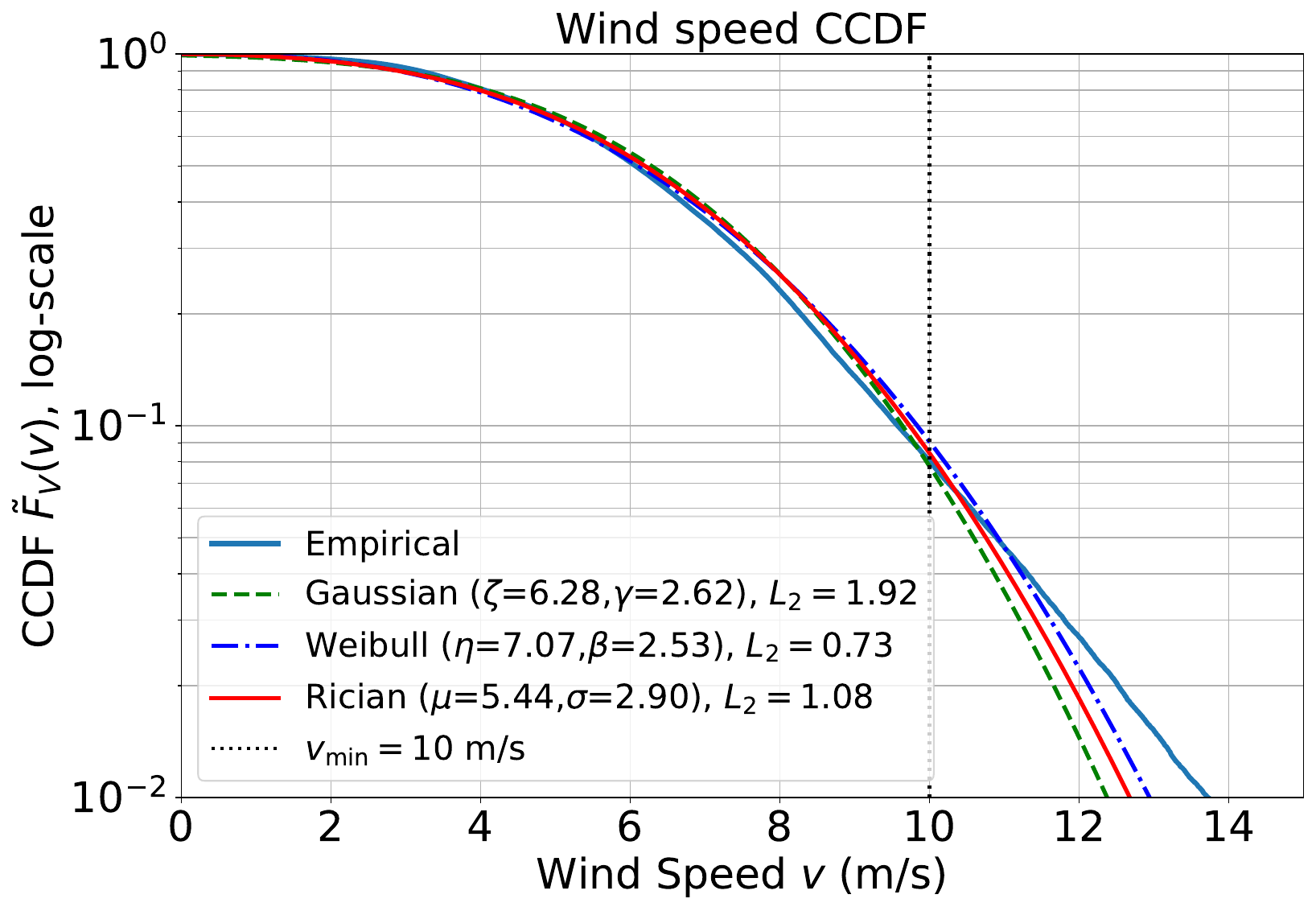}
         \caption{}
         \label{fig:ccdf_plot}
     \end{subfigure}
     
     \caption{Empirical wind-speed (a) PDF, (b) CDF, and (c) CCDF for a representative turbine in Wind Farm~1, together with Gaussian, Weibull, and Rician model fits. The associated goodness-of-fit metrics (KL divergence, KS distance, and tail-restricted $L_2$ distance) are also shown.}
     \label{fig:wind_speed_fits}
\end{figure}

Figure~\ref{fig:pdf_plot} shows the empirical and fitted PDFs for Wind Farm~1. Across all farms (Table~\ref{tab:wind_farm_comparison}), KL-divergence values are generally comparable among the three models, with the Gaussian fit typically showing slightly larger deviations in the bulk distribution.

The cumulative distribution function (CDF), $F_V(v_0)$ and complementary cumulative distribution function (CCDF), $\tilde{F}_V(v_0)$ \cite{book_majumdar,sabhapandit2019extremesrecords,book_bala} are defined as:

\begin{equation}\label{CDF_speed}
F_V(v_0) = \mathbb{P}(V \le v_0)
\end{equation}
\begin{equation}\label{CCDF_speed}
\tilde{F}_V(v_0) = \mathbb{P}(V > v_0) = 1 - F_V(v_0)
\end{equation}

For CDF-based comparison, we compute the Kolmogorov--Smirnov (KS) distance \cite{Newman_2005,Alstott_2014},

\begin{equation}
\label{ks}
D = \sup_{v_{i}} \left| F_{\mathrm{empirical}}(v_i) - G_{\mathrm{model}}(v_i) \right|
\end{equation}
where $F_{\mathrm{empirical}}$ and $G_{\mathrm{model}}$ are the empirical and fitted CDFs, respectively. Figure~\ref{fig:cdf_plot} presents the CDF comparison for Wind Farm~1, and KS values for all farms are reported in Table~\ref{tab:wind_farm_comparison}. The Gaussian model tends to perform worst in the low-speed regime, whereas Weibull and Rician models better capture the low-speed regime.

For high-wind events, we focus on the CCDF tail (Fig.~\ref{fig:ccdf_plot})\cite{book_majumdar,sabhapandit2019extremesrecords}. Because global metrics such as KL divergence and KS distance are not specifically tail-sensitive, we additionally use a tail-restricted logarithmic $L_2$ metric for $v \ge v_{\min}=10\,\mathrm{m\,s^{-1}}$:

\begin{equation}\label{l2_dist_tail}
L_2 = \frac{1}{N_{\mathrm{tail}}}
\sum_{v_i \ge v_{\min}}
\left[\log \tilde{F}_{\mathrm{empirical}}(v_i) -
\log \tilde{G}_{\mathrm{model}}(v_i)\right]^2
\end{equation}
Here, $N_{\mathrm{tail}}$ is the number of observations in the tail region, and $\tilde{F}_{\mathrm{empirical}}$ and $\tilde{G}_{\mathrm{model}}$ are the empirical and fitted CCDFs. Tail-$L_2$ values for all models and farms are summarized in Table~\ref{tab:wind_farm_comparison}.

Table~\ref{tab:wind_farm_comparison} indicates that the Gaussian model under-performs in the statistical tails. In that region, the Gaussian distribution decays rapidly as $f_G(v)\sim \exp(-v^2)$, unable to capture the large wind events. In contrast, the Rician distribution exhibits a slower decay, $f_R(v)\sim \sqrt{v}\exp(-v^2)$, while the Weibull tail behavior depends sensitively on the shape parameter $\beta$.

\begin{figure}[htb!]
\centering
\includegraphics[width=1\linewidth]{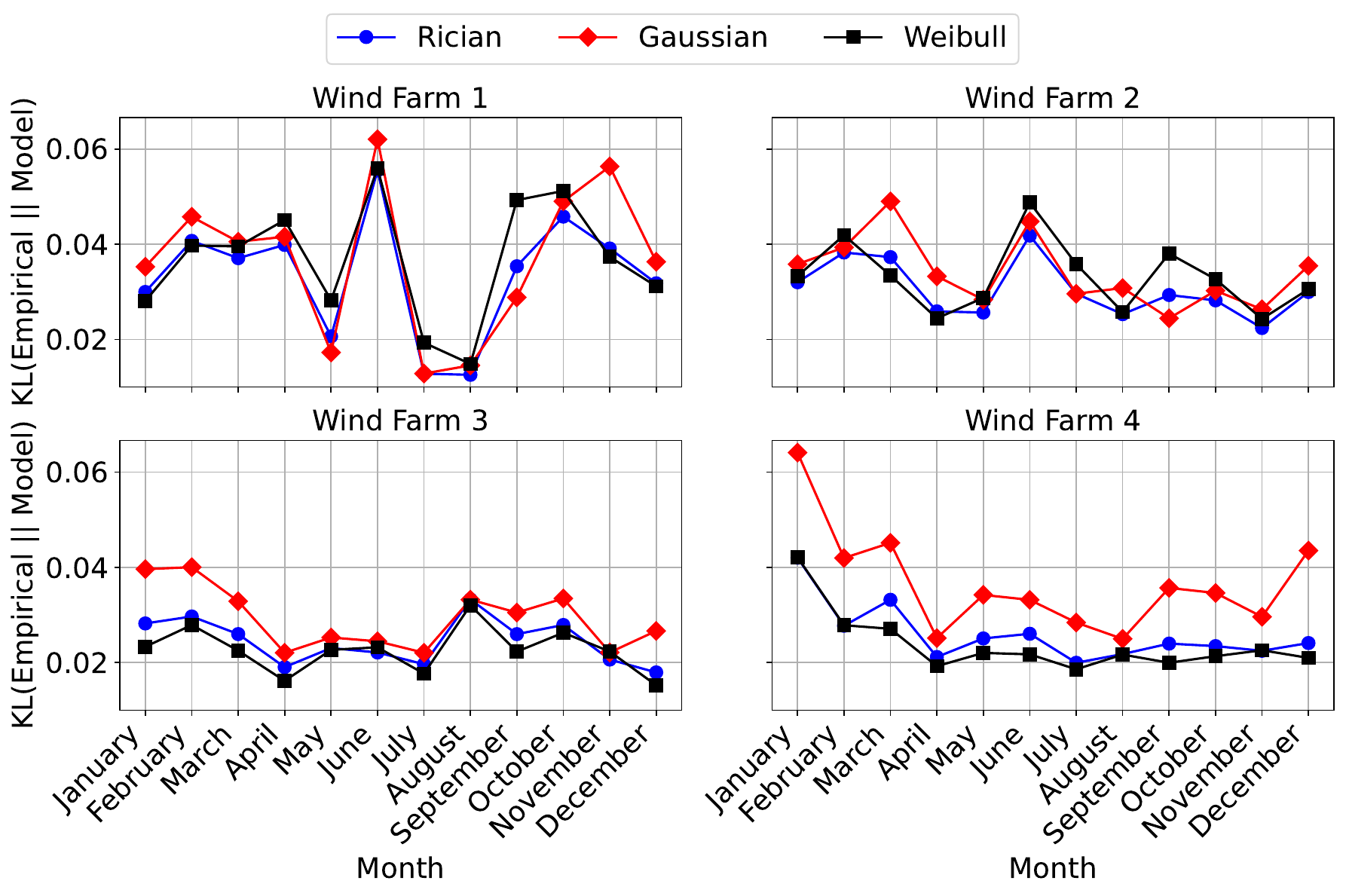}
\caption{Monthly averaged KL-divergence values for Gaussian, Weibull, and Rician fits across the four wind farms (averaged over all available years for each calendar month).}
\label{fig:kl_div_monthly}
\end{figure}

\begin{figure}[htb!]
\centering
\includegraphics[width=1\linewidth]{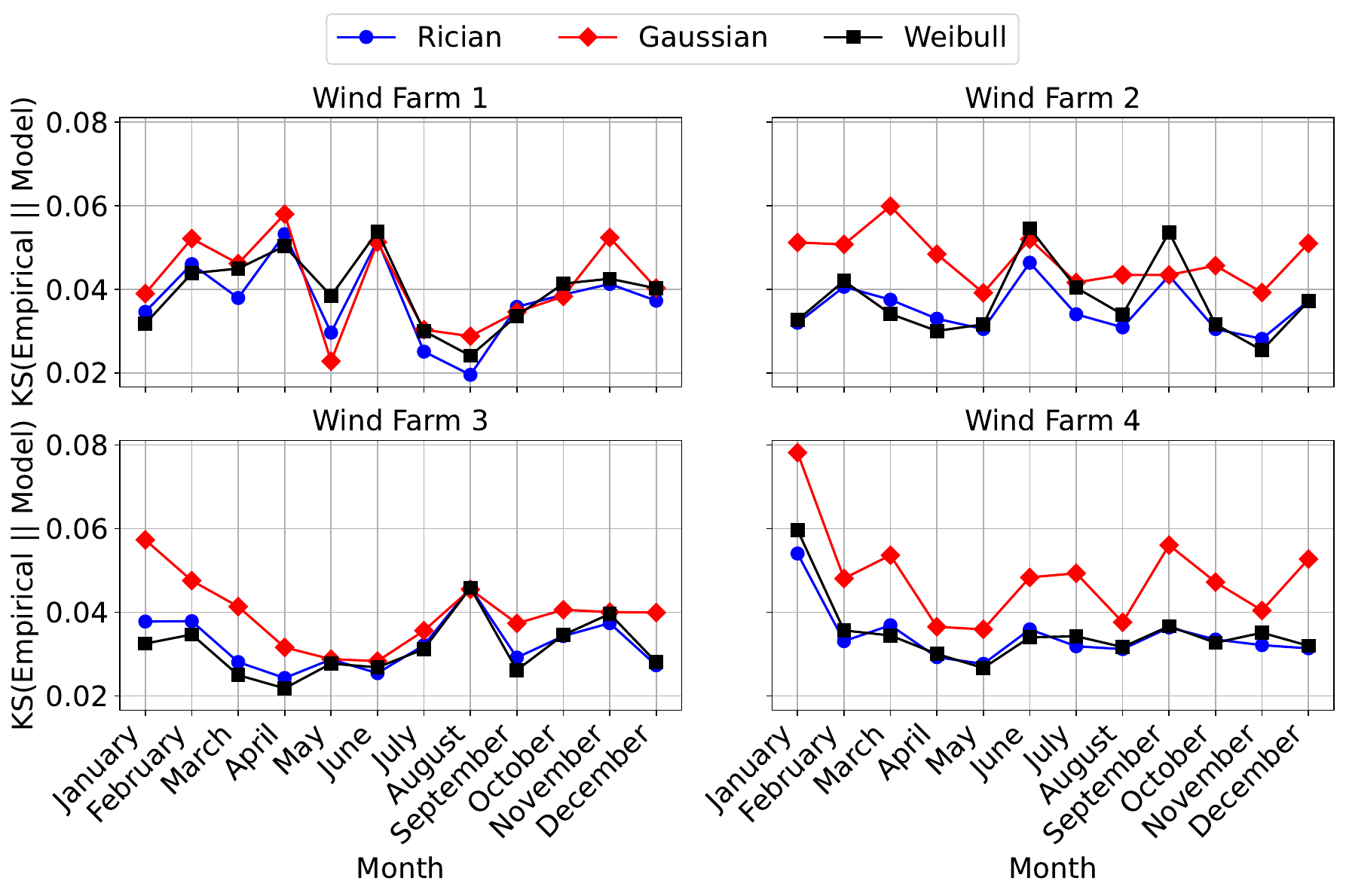}
\caption{Monthly averaged Kolmogorov--Smirnov (KS) distances for Gaussian, Weibull, and Rician fits across the four wind farms (averaged over all available years for each calendar month).}
\label{fig:ks_score_monthly}
\end{figure}

Motivated by the seasonal variations induced nonstationarity of the wind speed, we further evaluate model performance on monthly windows in addition to full-record fits. This analysis tests whether the relative model performance remains stable across changing seasonal wind regimes. We focus on two global metrics, KL divergence and KS distance, computed month by month for each wind farm. Figures~\ref{fig:kl_div_monthly} and~\ref{fig:ks_score_monthly} report the monthly averaged scores (averaged over all available years (see Table~\ref{Table:1}) for each calendar month), for each farm. Across all farms, the Gaussian model consistently yields larger KL and KS values, indicating poorer agreement with the empirical distributions. The Rician and Weibull models generally show comparable and substantially improved goodness-of-fit.

In summary, the Rician model consistently outperforms the Gaussian model and achieves goodness-of-fit comparable to the Weibull model, both for full-record fits and for monthly-window analysis. 

\section{Seasonal and Geographic Variability of Model Parameters \label{sec:seasonal_geo_variability}}

To evaluate physical interpretability, we examine how the fitted Rician distribution parameters vary by season and by site across the four wind farms. For each turbine and month, $\mu$ and $\sigma$ are estimated from monthly windows using MLE [Eq.~\ref{mle}] with the Rician model in Eq.~\ref{rician_pdf}. We then compute farm-level ensemble means [Eq.~\ref{farm_avg}] and inter-turbine spreads [Eq.~\ref{farm_std}] for each parameter $p\in\{\mu,\sigma\}$.

\begin{figure}[htb!]
\centering
\includegraphics[width=1\linewidth]{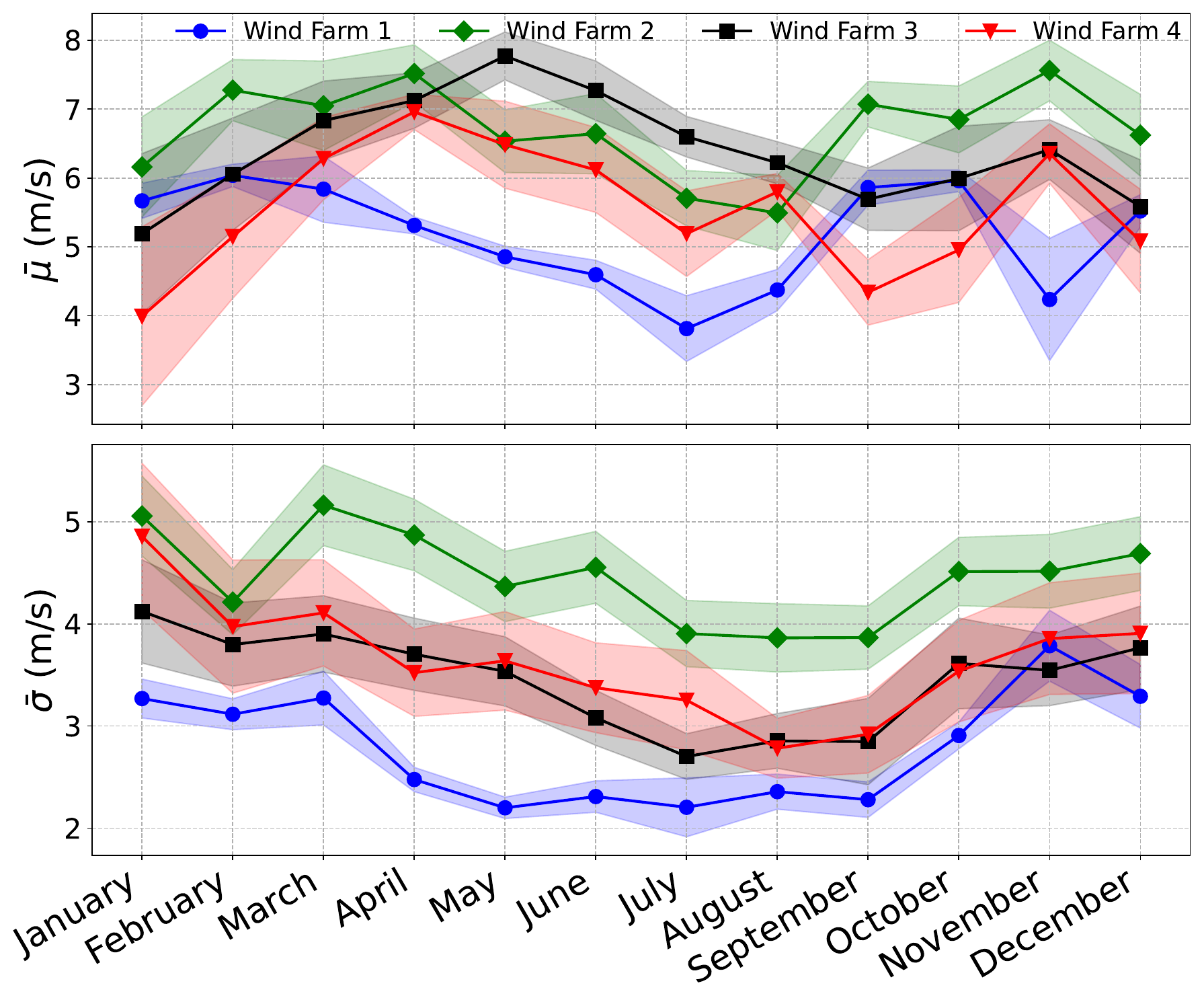}
\caption{Turbine ensemble averaged Rician parameters $\bar{\mu}$ and $\bar{\sigma}$ for the four wind farms, averaged over all available years for the four wind farms (see Table~\ref{Table:1}). Solid lines indicate farm-level ensemble average across all turbines [Eq.~\ref{farm_avg}]; shaded regions indicate inter-turbine variability [Eq.~\ref{farm_std}].}
\label{fig:rician_params}
\end{figure}

In Fig.~\ref{fig:rician_params}, bold curves denote farm-level ensemble-averaged values $\bar{p}_m$  [Eq.~\ref{farm_avg}], and shaded bands show inter-turbine variability up to one standard deviation, $\bar{p}_m \pm s_{p,m}$ [Eq.~\ref{farm_std}], as a function of month. The bands remain moderate but non-negligible, indicating that turbines within the same farm share common seasonal forcing while still experiencing distinct local gust conditions. The fitted parameters therefore capture both inter-farm and intra-farm spatial variability.

The same Figure.~\ref{fig:rician_params} \textcolor{gray}{also} shows that the parameters also capture temporal variability in a physically interpretable way. In particular, for Wind Farm~1 (Fig.~\ref{fig:monthly_timeseries}), the seasonal evolution of $\mu$ is consistent with the monthly mean wind-speed trend $\mu_E$, while $\sigma$ follows the seasonal modulation of fluctuation intensity $\sigma_E$, with similar kind of inter-turbine variability band.

Overall, these results support the Rician model as a compact and physically meaningful representation of wind-speed variability: it retains competitive goodness-of-fit, while providing parameters that directly characterize mean flow and stochastic variability across seasons and sites.

\section{Connection Between the Three Wind-Speed Distributions}
\label{sec:model_comparisons}

More generally, the Rician distribution is useful because it belongs to a broader class of models that interpolate between qualitatively different statistical regimes through a small number of control parameters. In particular, it provides a simple and physically interpretable bridge between Gaussian-like behavior, dominated by a coherent mean, and Weibull-like behavior, dominated by fluctuations \cite{Yang,9397402}.

We finally examine the limits in which the Rician model [Eq.~\ref{rician_pdf}] recovers the Gaussian [Eq.~\ref{gaussian_pdf}] and Weibull [Eq.~\ref{Weibull_pdf}] families. The shape of the Rician distribution is controlled by the dimensionless ratio $\frac{\mu}{\sigma}$ \cite{Yang,9397402}. A complete derivation is provided in \hyperref[appendix2]{Appendix B}.

Starting from Eq.~\ref{rician_pdf}, in the regime $v\approx\mu\gg\sigma$ (i.e. instantaneous speed close to the mean parallel speed and weak fluctuations), the Rician distribution approaches a Gaussian form [Eq.~\ref{ric_gau}]. In this limit, the Kullback--Leibler (KL) divergence in Eq.~\ref{kl_div} between Rician and Gaussian becomes minimal. Empirically, the two become nearly indistinguishable for $\frac{\mu}{\sigma}>3$.

In the opposite regime, $\sigma\gg\mu$, the Rician distribution approaches a Weibull-type form [Eq.~\ref{eq-WeibullExpansion}] with effective parameters $\beta\approx2$ and $\eta=\sqrt{2}\,\sigma$. In this limit, the KL divergence [Eq.~\ref{kl_div}] between Rician and Weibull is minimal. The special case $\mu=0$ reduces exactly to the Rayleigh distribution [Eq.~\ref{rayleigh}].

\begin{figure}[htb!]
\centering
\includegraphics[width=1\linewidth]{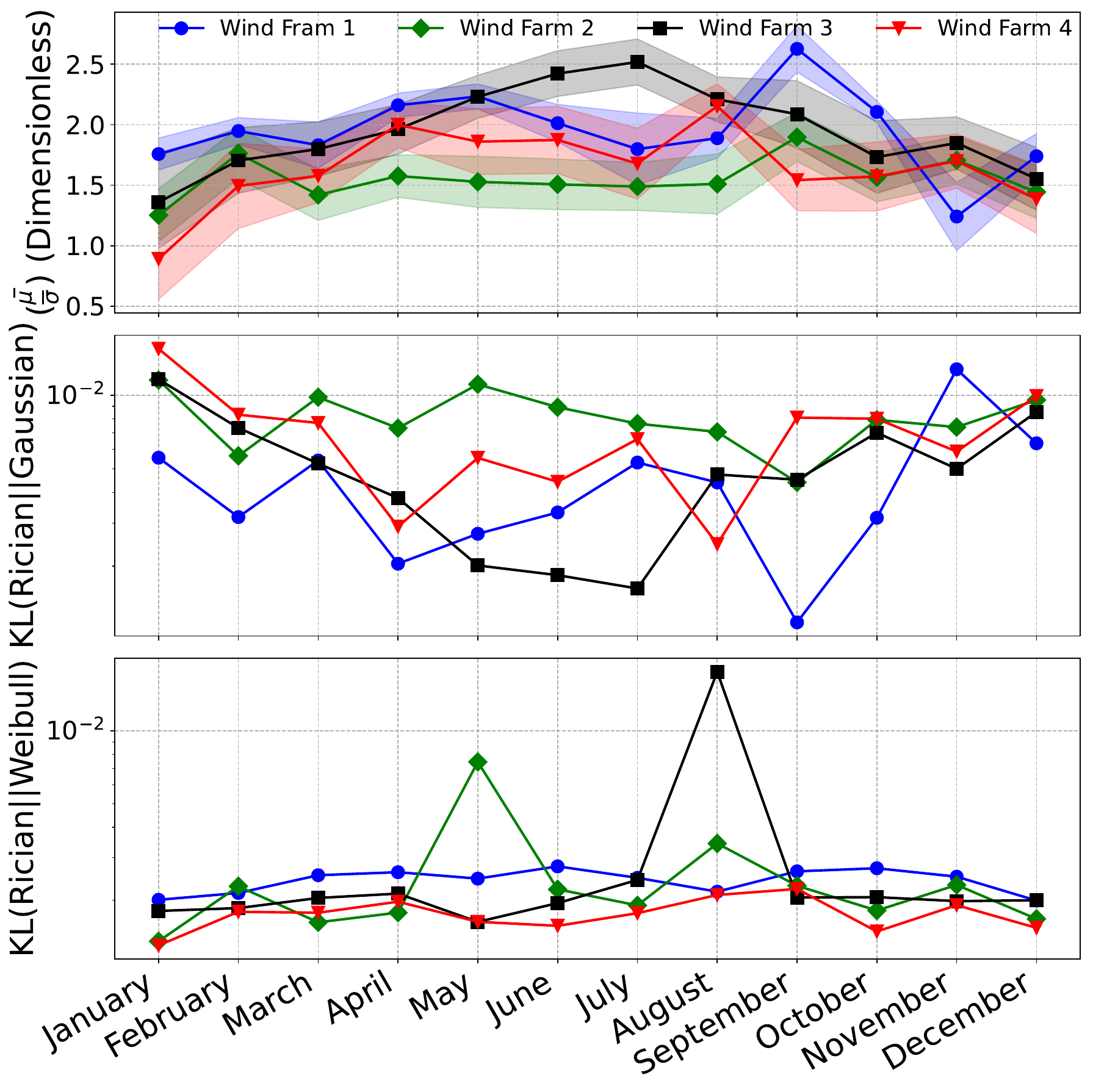}
\caption{Farm-averaged ratio $\frac{\mu}{\sigma}$ and corresponding KL-divergence trends quantifying the Rician-to-Gaussian and Rician-to-Weibull limits across the four wind farms.}
\label{fig:rician_ratio}
\end{figure}

Figure~\ref{fig:rician_ratio} summarizes this transition behavior. The first panel shows the farm-averaged ratio $\bar{(\frac{\mu}{\sigma})}$ (computed from Eq.~\ref{farm_avg}) together with inter-turbine spread (from Eq.~\ref{farm_std}, shown as shaded bands). The remaining panels show farm-averaged KL divergences between Rician and Gaussian, and between Rician and Weibull.

A consistent trend is observed: when $\bar{(\frac{\mu}{\sigma})}$ increases, the KL divergence between Rician and Gaussian decreases (Rician becomes more Gaussian-like), while the KL divergence between Rician and Weibull increases. The opposite trend occurs when $\bar{(\frac{\mu}{\sigma})}$ decreases. Wind Farms 1 and 3 generally exhibit larger $\bar{(\frac{\mu}{\sigma})}$ values (with seasonal modulation), and therefore lower Rician--Gaussian KL divergence and higher Rician--Weibull KL divergence than the other farms.

This indicates that the Rician model consistently interpolates between the canonical Gaussian and Weibull limits, both from theoretical considerations and in empirical wind-speed data across seasonal and geographic variations. As wind conditions evolve, the unified Rician framework naturally recovers Gaussian-like or Weibull-like wind-speed statistics in the appropriate regimes.

\section{Conclusions}
\label{sec:conclusions}

In this work, we proposed and validated the Rician distribution [Eq.~\ref{rician_pdf}] as a minimal model for wind-speed statistics. The Rician model provides a physically grounded description of wind-speed variability using only two parameters (Fig.~\ref{fig:rician_params}). We tested this framework using multi-year, 10-minute averaged measurements from four geographically distinct utility-scale wind farms (Table~\ref{Table:1}) and compared its performance against the canonical Gaussian [Eq.~\ref{gaussian_pdf}] and Weibull [Eq.~\ref{Weibull_pdf}] models.

Our results show that the Rician model successfully captures the main statistical features of empirical wind-speed distributions (Fig.~\ref{fig:monthly_hist}), including both the bulk of the distribution and the heavy upper tail (Fig.~\ref{fig:wind_speed_fits}). Across all wind farms, it consistently improves upon the Gaussian model and remains broadly comparable to the Weibull distribution (Table.~\ref{tab:wind_farm_comparison}) in terms of global goodness-of-fit measures such as the KL divergence (Fig.~\ref{fig:pdf_plot}, Fig.~\ref{fig:kl_div_monthly}) and KS distance (Fig.~\ref{fig:cdf_plot}, Fig.~\ref{fig:ks_score_monthly}). At the same time, it achieves competitive tail performance as quantified by the CCDF-based $L_2$ metric (Fig.~\ref{fig:ccdf_plot}). These results demonstrate that the Rician distribution provides a parsimonious yet accurate description of wind-speed statistics, with direct relevance for wind-resource assessment, forecasting, and reliability-oriented power system analyses.

As a natural next step, this framework can be extended from wind-speed statistics to wind-power statistics at the turbine level. Because the kinetic energy flux of the wind scales with the cube of the wind speed, turbine power output depends nonlinearly on wind-speed fluctuations \cite{bel2016grid,bandi2017spectrum}. Understanding how the Rician wind-speed statistics propagate through this nonlinear transformation may provide improved models for wind-power variability and extremes. This connection between wind-speed statistics and power-output variability may offer a simple yet physically interpretable framework for modeling renewable power fluctuations in large-scale power systems.

\appendix
\section{Appendix A: Derivation of the Rician distribution \label{appendix1}}

To derive the Rician speed distribution, we assume the parallel $v_{\parallel}$ and the perpendicular component  $v_{\perp}$, of the wind velocity  $\vec{v}$ at a fixed point in space, are distributed independently as Gaussian distributions:

\begin{equation}
\label{gaussian_decom}
\begin{aligned}
f_{\parallel}(v_{\parallel})
&= \frac{1}{\sqrt{2\pi\sigma^2}}
   \exp\!\left[-\frac{(v_{\parallel}-\mu)^2}{2\sigma^2}\right] \\
f_{\perp}(v_{\perp})
&= \frac{1}{\sqrt{2\pi\sigma^2}}
   \exp\!\left[-\frac{v_{\perp}^2}{2\sigma^2}\right]
\end{aligned}
\end{equation}

The wind speed is given by
\begin{equation}
v = \sqrt{v_{\parallel}^2 + v_{\perp}^2}
\end{equation}

Where the supports are $-\infty \leq v_{\parallel} \leq +\infty$ and $-\infty \leq v_{\perp} \leq +\infty$ imply that $0 \leq v \leq +\infty$.

The distribution of the wind speed $v$, $f_V(v)$, can then be written as:

\begin{multline}
f_V(v)
  =\int dv_{\parallel} \int dv_{\perp} f_{\parallel}(v_{\parallel})f_{\perp}(v_{\perp}) \delta(\sqrt{v_{\parallel}^2 + v_{\perp}^2}-v) \\
  = \frac{1}{2\pi\sigma^2}
    \int_{-\infty}^{+\infty}\!dv_{\parallel}
    \int_{-\infty}^{+\infty}\!dv_{\perp}\,
    \exp\!\left[-\frac{(v_{\parallel}-\mu)^2+v_{\perp}^2}{2\sigma^2}\right] \\
    \times \delta(\sqrt{v_{\parallel}^2 + v_{\perp}^2}-v)
\end{multline}

Using polar coordinates $v_{\parallel}= r\cos\theta$ and $v_{\perp} = r\sin\theta$, such that $v_{\parallel}^2 + v_{\perp}^2 = r^2$ and $dv_{\parallel}dv_{\perp} = rdrd\theta$, with limits $r \in [0, +\infty)$ and $\theta \in [0, 2\pi]$, we obtain

\begin{multline}
f_V(v) 
   =\frac{1}{2\pi\sigma^2} \int_{0}^{+\infty}dr\int_{0}^{2\pi}d\theta \\
   \times \exp[-\frac{(rcos\theta-\mu)^2+r^2sin^2\theta} {2\sigma^2}] r \delta(r-v)
\end{multline}

Performing the radial integration and simplifying gives

\begin{equation}\label{theta_integral}
f_V(v) =  \frac{v}{2\pi\sigma^2}
\exp\left[-\frac{v^2+\mu^2}{2\sigma^2}\right]
\int_{0}^{2\pi}d\theta\exp\left[\frac{\mu v\cos\theta}{\sigma^2}\right]
\end{equation}

Using the integral representation of the modified Bessel function of the first kind of order $l$ \cite{book_bala}:

\begin{equation}
I_l(z)=\frac{1}{\pi}\int_{0}^{\pi}d\theta e^{z\cos\theta}\cos(l\theta)
\end{equation}

Substituting $I_0$ into Eq.~\ref{theta_integral}, yields the final form of the wind speed distribution, known as the Rician or Rice distribution \cite{article}:

\begin{equation}
f_R(v;\mu,\sigma) = \frac{v}{\sigma^2}
\exp\left(-\frac{v^2+\mu^2}{2\sigma^2}\right)
I_0\left(\frac{v\mu}{\sigma^2}\right), \quad v \geq 0
\end{equation}

\section{Appendix B: Transition conditions from Rician to Gaussian and Weibull distributions \label{appendix2}}

To derive the Rician-to-Weibull and Rician-to-Gaussian limits, we first rewrite the Rician PDF [Eq.~\ref{rician_pdf}] in nondimensional form.
Define
\begin{align*}
z &= \frac{v}{\sigma}, \\
\lambda &= \frac{\mu}{\sigma},
\end{align*}
with $z\ge 0$ and $\lambda\ge 0$. Then
\begin{equation}
\label{eq-Rician}
f_R(z;\lambda)= z\,\exp\!\left[-\frac{z^2+\lambda^2}{2}\right]I_0(\lambda z)
\end{equation}

\subsection{Weibull (Rayleigh) limit: $\lambda\ll 1$}

For small $\lambda$, expand each factor in Eq.~\ref{eq-Rician}  \cite{book_bala}:
\begin{align*}
\exp\!\left(-\frac{\lambda^2}{2}\right) &= 1-\frac{\lambda^2}{2}+\mathcal{O}(\lambda^4)\\
I_0(\lambda z) &= 1+\frac{\lambda^2 z^2}{4}+\mathcal{O}(\lambda^4)
\end{align*}
Multiplying,
\begin{equation}
\label{eq-WeibullExpansion}
f_R(z;\lambda)= z\,e^{-z^2/2}\left[1+\frac{\lambda^2}{4}(z^2-2)+\mathcal{O}(\lambda^4)\right]
\end{equation}
The leading term is exactly the Rayleigh PDF in $z$,
\begin{equation}
\label{rayleigh}
f_{\mathrm{Ray}}(z)=z e^{-z^2/2}
\end{equation}
which is a Weibull distribution with shape parameter $\beta=2$ and scale parameter $\eta=\sqrt{2} \, \sigma$.
Hence, as $\lambda=\frac{\mu}{\sigma}\to 0$, the Rician distribution converges to the Weibull limit (specifically, the Rayleigh case \cite{WADI2023237}).

A uniform validity condition for the truncated expansion is
\begin{equation}
\left|\frac{\lambda^2}{4}(z^2-2)\right|\ll 1
\end{equation}
which is the precise version of the small-correction requirement.

\subsection{Quasi-Gaussian and Gaussian limits: $\lambda\gg 1$}

Rewrite Eq.~\ref{eq-Rician} as
\begin{equation}
\label{ric_weib}
f_R(z;\lambda)= z\,e^{-(z-\lambda)^2/2}F(\lambda z)
\end{equation}
where
\begin{equation}
F(x)=e^{-x}I_0(x)
\end{equation}
Using the asymptotic expansion \cite{book_bala}, as $x\to\infty$
\begin{equation}
F(x)=\frac{1}{\sqrt{2\pi x}}\left[1+\frac{1}{8x}+\mathcal{O}\!\left(\frac{1}{x^2}\right)\right]
\end{equation}
Substituting $x=\lambda z$ gives
\begin{equation}
\label{eq-QGExactAsympt}
f_R(z;\lambda)=\sqrt{\frac{z}{2\pi\lambda}}\,e^{-(z-\lambda)^2/2}
\left[1+\frac{1}{8\lambda z}+\mathcal{O}\!\left(\frac{1}{(\lambda z)^2}\right)\right]
\end{equation}
Dropping higher-order terms yields the quasi-Gaussian approximation,
\begin{equation}
\label{eq-QGApprox}
f_R^{\mathrm{QG}}(z;\lambda)\approx\sqrt{\frac{z}{2\pi\lambda}}\,e^{-(z-\lambda)^2/2}
\left(1+\frac{1}{8\lambda z}\right)
\end{equation}
valid when $\lambda z\gg 1$ (in practice, away from $z=0$).
To obtain a pure Gaussian core, set $z=\lambda+\delta$ with $|\delta|\ll\lambda$. Then
\begin{equation}
\sqrt{\frac{z}{\lambda}}=\sqrt{1+\frac{\delta}{\lambda}}=1+\mathcal{O}\!\left(\frac{\delta}{\lambda}\right)
\end{equation}
and
\begin{equation}
\frac{1}{\lambda z}=\frac{1}{\lambda^2}\left[1+\mathcal{O}\!\left(\frac{\delta}{\lambda}\right)\right]
\end{equation}
Therefore,

\begin{equation}
\label{ric_gau}
\begin{split}
f_R^{\mathrm{G}}(z;\lambda) &\approx \frac{1}{\sqrt{2\pi}}\,e^{-\delta^2/2} \left(1+\frac{1}{8\lambda^2}\right) \\
&= \frac{1}{\sqrt{2\pi}}\,e^{-(z-\lambda)^2/2}\left(1+\frac{1}{8\lambda^2}\right)
\end{split}
\end{equation}

Thus, for $\frac{\mu}{\sigma} \gg 1$, the Rician distribution is asymptotically Gaussian near its peak, with small $\mathcal{O}(\lambda^{-2})$ amplitude corrections.

In summary, small $\frac{\mu}{\sigma}$ gives the Weibull-like regime (Rayleigh limit), while large $\frac{\mu}{\sigma}$ gives a quasi-Gaussian distribution that reduces to a Gaussian in the near-peak region.

\begin{acknowledgments}
SL was supported by a Japan Society for the Promotion of Science (JSPS) Postdoctoral Fellowship (grant no. P24714). MB was supported by JSPS KAKENHI (Grant No. 24KF0079). The data used in this work was obtained from Scout Clean Energy, Boulder CO.
\end{acknowledgments}

\bibliography{pop}

\end{document}